# Spontaneous Charge aggregation and Polarization Model for Electric Discharges in Clouds

Maria Tsouchnika[1], Michael Kanetidis[1], Panos Argyrakis[1], and Raoul Kopelman[2,*]

Abstract:

In present models cloud based lightning forms as a consequence of the earth's gravitational and/or electromagnetic fields. Our simplified *field-free* model probes the random aggregation of a neutral ensemble consisting of a random distribution of equal numbers of equally charged positive and negative "frozen droplets", given a bivariate sticking probability. Over a wide range of parameters, one maximal cluster (called "cloud") is formed, with time-increasing mass, net charge and electric dipole moment, where all relations follow a simple argument, and the only parameter is the fractal dimension of the cloud: In this overall neutral system we find that the net charge of the cloud increases with the square root of the cloud's mass, while the electric dipole moment increases quasi-linearly and even superlinearly. *Thus an internal electric field can be self-created by a random process, with a field-strength that increases with time and cloud size, using just a two-dimensional random walk*. Potential implications of this toy model on "real life" clouds are discussed, i.e., the electric dipole moment based possibility of having intra-cloud lightning discharges, and the net electric charge based possibility of having extra-cloud lightning discharges, all in the absence of gravitational or other external fields or forces. Possible implications are discussed for both water and dust based clouds, atmospheric and extraterrestrial. Potential applications to biology are also discussed.

1. Introduction

The interest in the relation between electric charge formation and lightning goes back to antiquity but at least to Benjamin Franklin[1]. Clouds presumably form by the condensation and aggregation of water droplets, in various freezing states. Each "droplet" (could be an ice crystal or a "hailstone") may contain a minute amount of charge, negative or positive. Though a complete picture may not yet have been established, the "standard model" assumes gravity induced height segregation between the positive and negative charge carrying units[2]. Also, the lightning discharge to ground is attributed to the latter's surface charge[3-4].

For the aggregating "cloud" we assume an equal probability of positive or negative charges (of equal magnitude) in the form of electrically insulating single frozen "droplets". Depending on whether two neighboring droplets have equal or opposite charges, there is a minor Coulomb repulsion or attraction between the two, respectively. We thus look at the "cloud" formation as a random aggregation of droplets, with a bias for nearest neighbors to be of opposite charge. The question of interest is: **How can such a process generate an electric field in the cloud, or, alternatively, form a cloud-wide large electrical dipole?** Again, in our model we simply assume that each droplet can be represented by a fixed absolute charge, positive or negative, and that the sticking probability between any two depends on their positive, or negative attraction. The cloud formation process is modeled by the formation of clusters, followed by the aggregation of such clusters into progressively larger ones, all based on a simple random walk and sticking probability model. A well-known similar simple model, Diffusion Limited Aggregation (DLA)[5] has found much application in physics, chemistry and biology[6].

Can our model be applied to atmospheric clouds? Notably, additional parameters, such as gravity, are excluded from our model. Could it also apply to interstellar clouds? Also, a conundrum similar to that of storm cloud formation (but with no lightning and thunder) appears to be the unexpected formation of very large electric fields in the cytosol of a biological cell[7,8]. Presumably there their formation has to do with the aggregation of proteins, with their inherent charged groups. While this protein aggregation may be of a more complicated nature, involving not only translation and sticking, but also protein orientation (and possible conformational changes and/or hydrogen bonds), the simple model presented here for the process of storm cloud formation may serve as a very rough first approximation for the cellular process, or may at least serve as a guide for forming a reasonable model that represents the biological situation.

In previous work[9,10,11] the distributions of the sizes of clusters formed by such aggregates were studied both for 2-D and 3-D lattices, and the corresponding dynamic scaling functions were derived. In these works no charges were included, but all clusters were made of neutral particles. Aggregation of oppositely charged particles were reported[12] for experiments with polystyrene spheres, in which several clusters were created with different structure, which depended strongly on the background electrolyte. Several other works have been reported in the literature[13,14,15,16,17], mainly on gelation, deriving models of cluster aggregation, and deriving the resulting fractal dimensions.

All the above works were concerned only with the structural characteristics of the ensuing aggregate but not the resulting net charge and dipole moment. In this paper we present results from simulations of cluster aggregation made of positively and negatively charged particles (droplets), which are initially randomly distributed, they diffuse with a random-walk type model, and they coalesce upon collision with different probabilities for like-like and like-unlike particles. We find that the resulting structures depend strongly on the initial particle concentration, as exemplified by the calculated fractal dimension. For low particle concentrations the fractal dimension is close to 1, resulting in a linear type of cluster (see example in Fig.1), while for higher concentrations the resulting clusters are more compact with fractal dimensions up to 1.75. Our next result shows that the accumulated charge on the largest cluster increases with time and cluster mass. As a consequence of this, we find that an electric dipole moment is generated, which also increases with time and cluster mass.

## 2. Method of Simulation

Positively charged, A, and negatively charged, B droplets are placed randomly on a two-dimensional lattice of size L x L, with density $d_A = d_B$, respectively. At this point the droplets are considered to be clusters of mass 1, with equal absolute charges. Clusters are selected at random and they perform a random walk on the lattice to the nearest neighbor sites, under the constraint that a lattice site cannot be occupied by more than one droplet at a time (excluded volume). When a move is to be made one of the four directions (up, down, left, and right) is chosen at random. If the chosen new position is not occupied the cluster moves into it, otherwise, it remains at its original position. All clusters have a mobility that is inversely proportional to their mass. This, in effect, means that a "lighter" cluster would move more frequently than a "heavier" one e.g. a cluster of mass 1 moves approximately in every MC step, whereas a cluster of mass 20 moves approximately once every 20 steps, etc. When all clusters have attempted to move once using this mechanism, regardless of whether the attempt is successful or not, this constitutes one Monte-Carlo time step (MCS). After all allowed moves have been made, any two clusters that happen to be now nearest neighbors are merged into one cluster, with probability p that depends on the type of the neighboring droplets of the clusters. The merging probability between two droplets of different type is $p_{A-B} = 1$, while the merging probability of two droplets of the same type is $p_{A-A} = p_{B-B}$ is an external parameter in the range $0<p<1$. The mass and charge of the derived cluster is the algebraic sum of the respective masses and charges of the formerly separate clusters. Periodic boundary conditions are applied on the lattice. We let the system evolve for several steps while monitoring the amount of absolute net charge and the dipole moment of the clusters formed. The process continues until all clusters have been merged to one final cluster. In Fig. 1 we show a pictorial for the process after 1 million steps. We give two different initial concentrations, in (1a) d=0.05, and (1b) d=0.15. We observe that in (1a), where the initial concentration is low, we see that a single cluster has been formed that is composed of positive and negative droplets, which is almost 1-dimesnional. In (1b), where the initial concentration is high, a more complex structure has been formed.

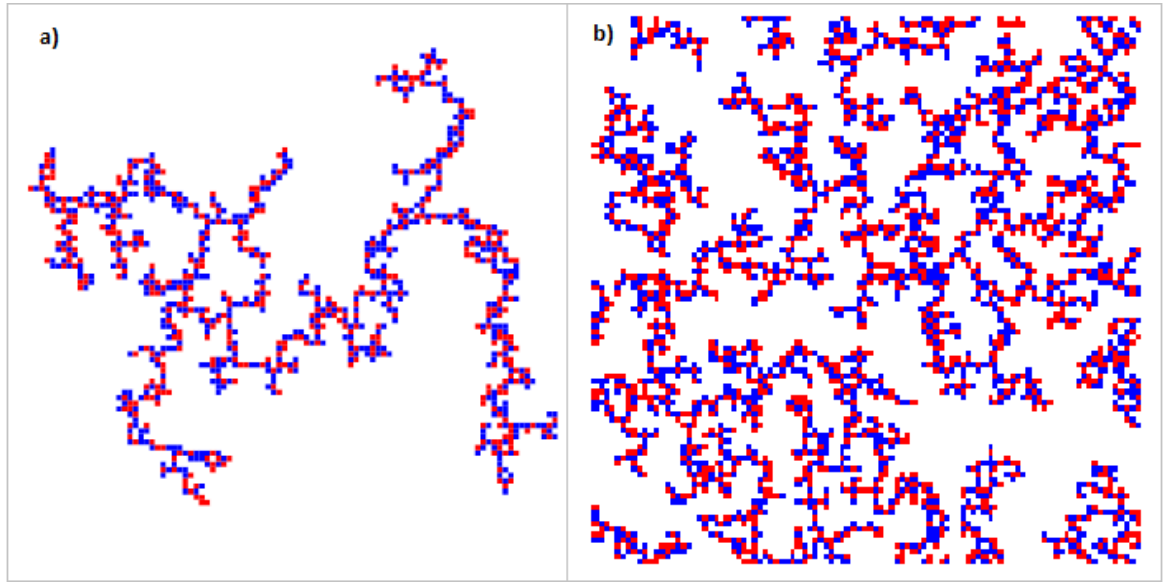

**Figure 1:** A pictorial of the process studied, after 1 million steps at which time a single cluster has been formed. Red and blue particles designate the two different charges. Density a) $d_A = d_B = 0.05$, b) $d_A = d_B = 0.15$, Lattice size: 100x100

3. Results

Charge Accumulation

We followed the evolution of the system for five different droplet densities, $d_A=d_B =$ 0.0025, 0.005, 0.025, 0.05 and 0.15. We allow the system to proceed to 300,000 MCS, and we isolate the largest (maximal) cluster that is formed. We plot the charge in this maximal cluster for these five densities, producing Fig. 2. Additionally, we do the same for the cluster that is carrying the maximal charge (Supplementary Fig. 5). All lines in Fig. 2 can be fitted as power-law curves, with approximately the same exponent for all densities examined. The calculated exponent (Fig. 2) is $a \cong 0.51$, as expected from the following argument. As the A and B droplets attach to a cluster with the same probability, the absolute net charge for a cluster consisting of a large number of droplets, n, is expected to be proportional to $n^{1/2}$. The same holds true for the line of Supplementary Fig. 5.

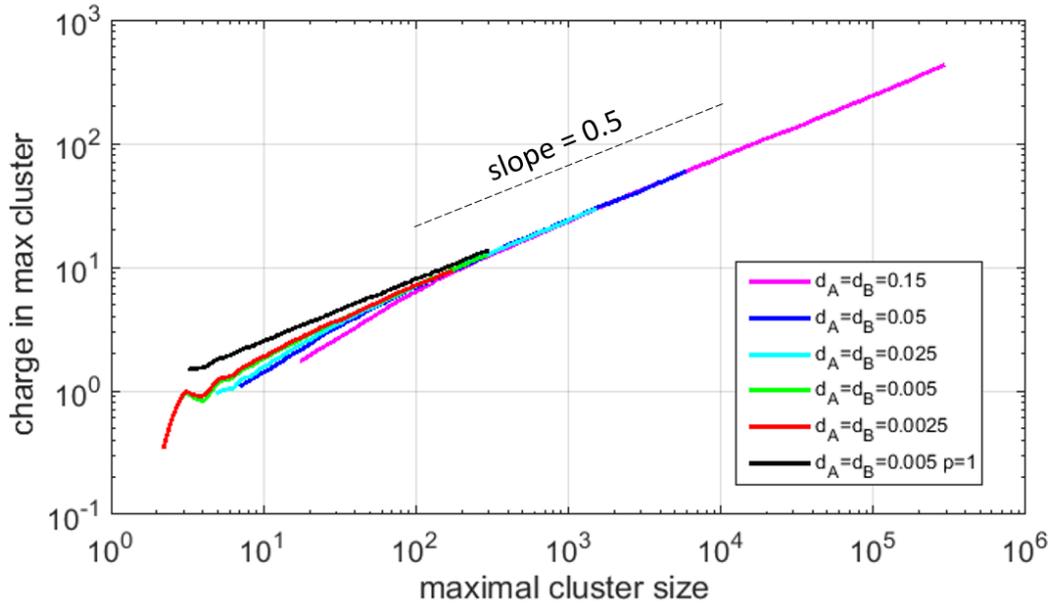

**Figure 2:** Charge in the largest cluster vs. its mass, in the case of a 1000x1000 lattice with periodic boundary conditions. Five droplet concentrations are studied in these systems after 300,000 Monte-Carlo steps. Averages are over 10,000 realizations.

Electric Dipole Moment

In Fig. 3 we calculate the dipole moment of the largest cluster as a function of its mass, for the same five different densities. We observe here that all curves for all densities collapse into one line with a slope α in the range ~1.07 – 1.24 and a mean value of approximately 1.15.

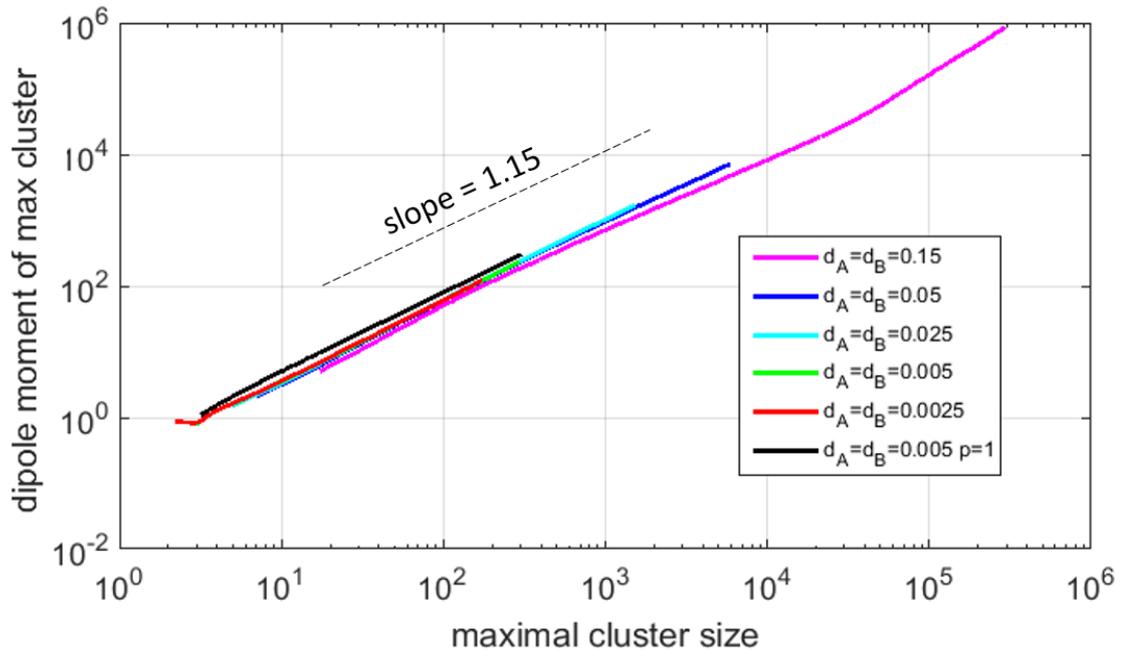

**Figure 3:** Dipole moment of largest cluster vs. its mass, in the case of a 1000x1000 lattice with periodic boundary conditions. Five droplet concentrations are studied in these systems after 300,000 Monte-Carlo steps. Average is over 10,000 realizations. The dotted line is a straight line with a slope =1.15.

Fractal Dimension

We calculated the fractal dimension $d_f$ of the maximal cluster using the box counting method[18], for several droplet concentrations. We plot our results in Fig. 4 (blue squares) and observe two different regimes. One for low droplet concentrations, where the rate of increment is slow, and one for higher concentrations, where the rate is higher, with a crossover at about density d≈0.03

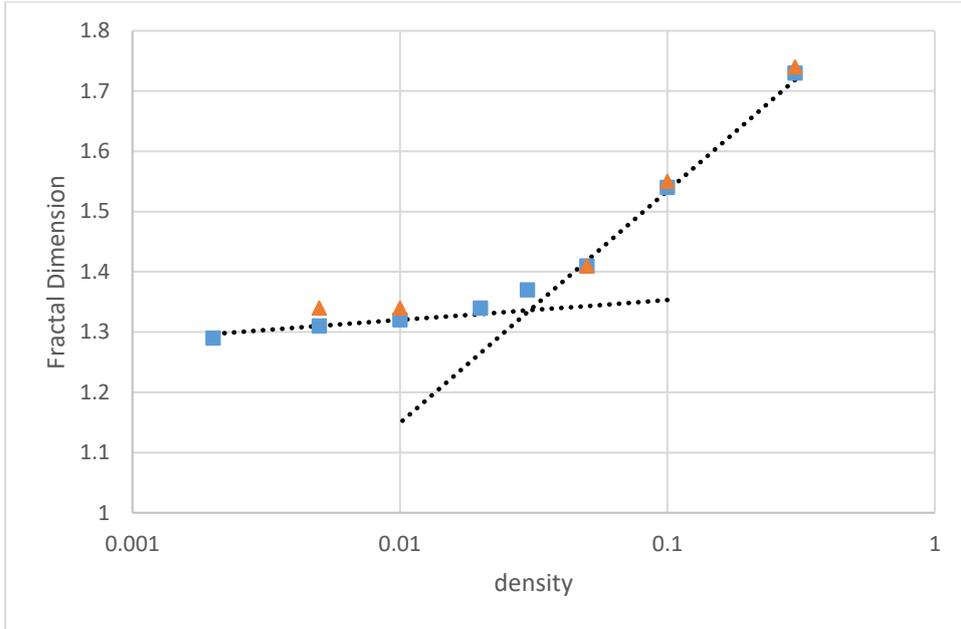

**Figure 4:** Fractal dimension of the largest cluster as a function of the total droplet concentration (blue squares). A number of 100 realizations were averaged. Red squares are the $d_f$ values as calculated by Eq. (5), see discussion later.

4. Discussion

Total Charge:

As the A and B droplets attach to a cluster with the same probability, the absolute charge for a cluster consisting of a large number of droplets, N, is expected to be proportional to $N^{1/2}$.

Let $q_i$ be the charge of droplet i, with $q_i$ = +1 or -1 with equal probability. Then for the total charge Q of a cluster of N droplets, we have:

$$Q = \sum_{i=1}^{N} q_i$$

$$Q^2 = \sum_{i=1}^{N} q_i \sum_{j=1}^{N} q_j = \sum_{i=1}^{N} (q_i)^2 + \sum_{i \neq j} q_i q_j = N + \sum_{i \neq j} q_i q_j$$

$$\langle Q^2 \rangle = N + \langle \sum_{i \neq j} q_i q_j \rangle = N$$

since $q_i$ = +1 or -1 with equal probability. Therefore,

$$\sqrt{\langle Q^2 \rangle} = \sqrt{N}$$
$$|Q| \sim \sqrt{N}$$

Charge to Mass Ratio:

Since the total charge increases with the cluster mass as Q ~ $N^{1/2}$ the charge to mass ratio decreases with the cluster mass as Q/N ~ $N^{1/2}$/N = $N^{-1/2}$.

Electric Dipole Moment

The dipole moment $\vec{p}$ of N charged particles and non-zero overall (total) charge is defined as:

$$\vec{p} = \sum (\vec{r}_i - \vec{r}_{ref}) \, q_i \tag{1}$$

where the sum runs over all charged particles of the cluster; the vector $\vec{r}_i$ is the position of the *i*th particle, measured from an arbitrary origin; $q_i$ is its charge and $\vec{r}_{ref}$ a reference point.

The results of the simulations indicate that the relation between the dipole moment P and the mass N of the cluster is (see Fig. 3):

$$P \sim N^a \tag{2}$$

where the exponent α is in the range 1.07 – 1.24.

This result is rationalized in the supplementary file, where we derive the following relation:

$$P \sim N^{\frac{1}{2} + \frac{1}{d_f}} \tag{3}$$

Therefore, the exponent α of Eq. (2) is:

$$\alpha = \frac{1}{d_f} + \frac{1}{2} \tag{4}$$

$$d_f = \frac{1}{a - \frac{1}{2}} \tag{5}$$

In Table 1 we show in the second column the slope as derived from the simulations. In the third column we list the fractal dimension of the maximal cluster, as it was found from the simulations using the box-counting method[18]. In the fourth column, we list the fractal dimension of the maximal cluster as it was calculated from Eq. (5), using the slope of the curves of Fig. 3 (second column).

| Total density | Slope ($\alpha$) from simulation | Fractal dim from simulation using box-count | Fractal dim from calculation (Eq. 5) |
|---|---|---|---|
| 0.3 | 1.073 | 1.73 | 1.74 |
| 0.1 | 1.145 | 1.54 | 1.55 |
| 0.05 | 1.210 | 1.41 | 1.408 |
| 0.01 | 1.245 | 1.32 | 1.342 |
| 0.005 | 1.242 | 1.31 | 1.348 |

**Table 1**: Fractal dimension of the maximal cluster from simulation using the box-counting method and calculations using Eq. (5)

Dipole moment to mass ratio:

Since the electric dipole moment increases with the cluster mass as $P \sim N^{1/2+1/d_f}$, the dipole moment to mass ratio also increases with the mass as $P/N \sim N^{-1/2+1/d_f}$, i.e. the exponent is positive for $d_f < 2$.

Fractal Dimension vs Concentration

In Fig. 4 we added (red triangles) the fractal dimension as calculated by Eq. (5), and we see that it is in very good agreement with the values derived from the box counting method (blue squares), with the same trends and crossover point as discussed earlier.

5. Conclusions

A simple hetero-aggregation model, starting with an equal number of insulator particles (frozen droplets) with positive and negative charges, of equal magnitudes, *randomly distributed in 2-dimensional space*, is able to show the formation of a *largest aggregate* ("cloud"), increasing in time, *with a net charge, as well as an electric dipole, also increasing in time,* while overall electrical neutrality is maintained. Regarding the growth in net charge, and dipole moment, with aggregate mass, the Monte-Carlo simulations agree with simple theoretical relations, where the fractal dimension of the aggregate mass is the only parameter. Notably, there are only internal Coulomb forces implied, but no gravity or external polarization fields, in contrast to existing models that explain cloud lightning phenomena. In principle, such a dipole could lead to an electric discharge or formation of "intra-cloud lightning", in the proper medium, with no external fields (such as gravity).

We believe that these considerations and results might be relevant to volcanic or interstellar clouds outside of Earth, e.g. on Mars, because particles are surrounded by a vacuum with no potential intercluster conduction. We also note that our model does not pertain to very high voltages that could lead to corona discharge mechanisms[4]. We know that lightning is initiated below the corona discharge, at electric fields of 250 kV/m or less[5].Thus we believe that we can ignore any effects due to Corona discharge below the initial discharge in the case of lightning.

Similarly, the net charge accumulations could explain inter-cloud and cloud-to-ground lightning discharge phenomena. It is intuitively obvious that, qualitatively, similar results would be gotten in 3-dimensions, with a quasilinear dependence of the dipole on the mass, and still a square root dependence of the charge. Several questions arise now as result of these findings. Could such a model be helpful in understanding natural or lab generated phenomena? Could it be relevant to atmospheric or even interstellar clouds? Could it be relevant to the build-up of electric fields in biological cells or subcellular compartments?

## Author information

**Affiliations**

1. **Department of Physics, Aristotle University of Thessaloniki, University Campus, 54124, Thessaloniki, Greece**
   M. Tsouchnika, M. Kanetidis, P. Argyrakis

2. **Department of Chemistry, University of Michigan, Ann Arbor, Michigan 48109 USA**
   R. Kopelman

**Contributions**

R.K. conceived the research. P.A. designed the Monte-Carlo model. M.T., M.K. performed the simulation calculations, and the theoretical analytical/scaling calculations. All authors discussed and related the physics models, wrote the paper, and reviewed the manuscript.

**Competing Interests**

The authors declare no competing interests.

**Corresponding author**

Correspondence to R. Kopelman, kopelman@umich.edu

# Spontaneous Charge aggregation and Polarization Model for Electric Discharges in Clouds

Maria Tsouchnika[1], Michael Kanetidis[1], Panos Argyrakis[1], and Raoul Kopelman[2,*]

[1] Department of Physics, University of Thessaloniki, 54124 Thessaloniki, Greece

[2] Department of Chemistry, University of Michigan, Ann Arbor, Michigan 48109 USA

*Corresponding Author:

Raoul Kopelman

Department of Chemistry, University of Michigan,

Ann Arbor, Michigan 48109 USA

kopelman@umich.edu

Number of Clusters

We monitor the number of clusters that are being formed as a function of time. We do this for five different initial droplet densities, $d_A = d_B$ = 0.0025, 0.005, 0.025, 0.05 and 0.15. The results are shown in Supplementary Fig. S1, where we observe that for concentrations d<0.05 the number of clusters converge to the same value for all densities, while for very dense systems with d ≥ 0.15 this number decreases very rapidly, approaching unity, i.e. a single cluster. We also note that for d=0.005 there is no difference in the number of clusters formed when p=1, i.e. if the charge plays no role at all. We also observe two different regimes, the one on the left where the cluster aggregation rate is very slow and after a characteristic time $t_d$ the second, where the curve seems to follow a power law. All results are averages over 10,000 realizations.

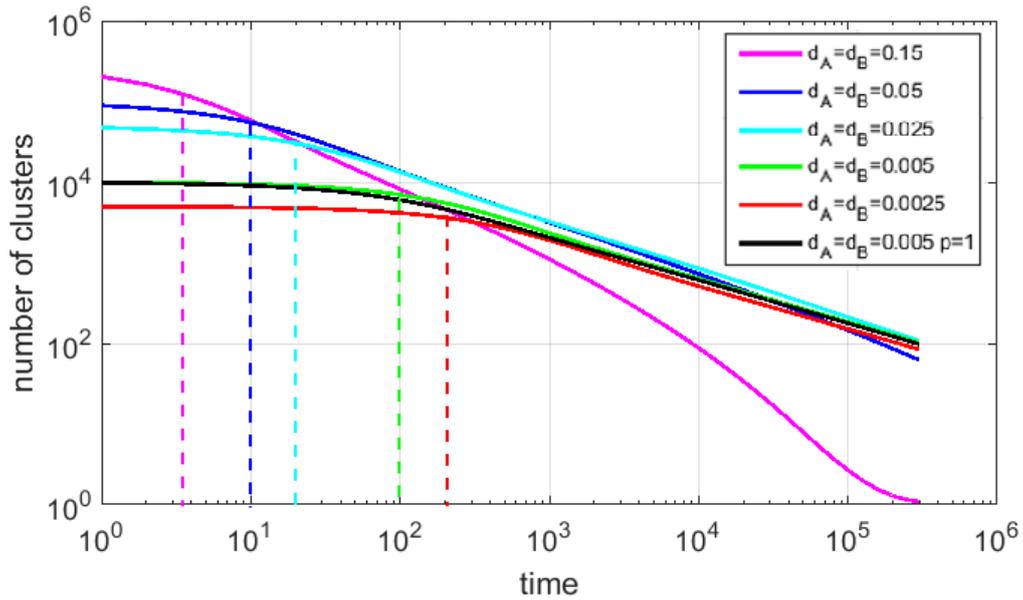

**Supplementary Figure S1:** Number of clusters vs. time, in the case of a 1000x1000 lattice with periodic boundary conditions and $d_A = d_B$. Five droplet concentrations are studied for 300,000 Monte-Carlo steps. Average is over 10,000 realizations. Vertical lines are drawn at the crossover times $t_c$ between the two regimes, for each concentration.

Mass of Largest Cluster

In Supplementary Fig. S2 we show the size of the maximal cluster as a function of time for several different droplet densities. We observe an initial slow increase of this size for times up to about 100 MCS, and a much faster increase at longer times.

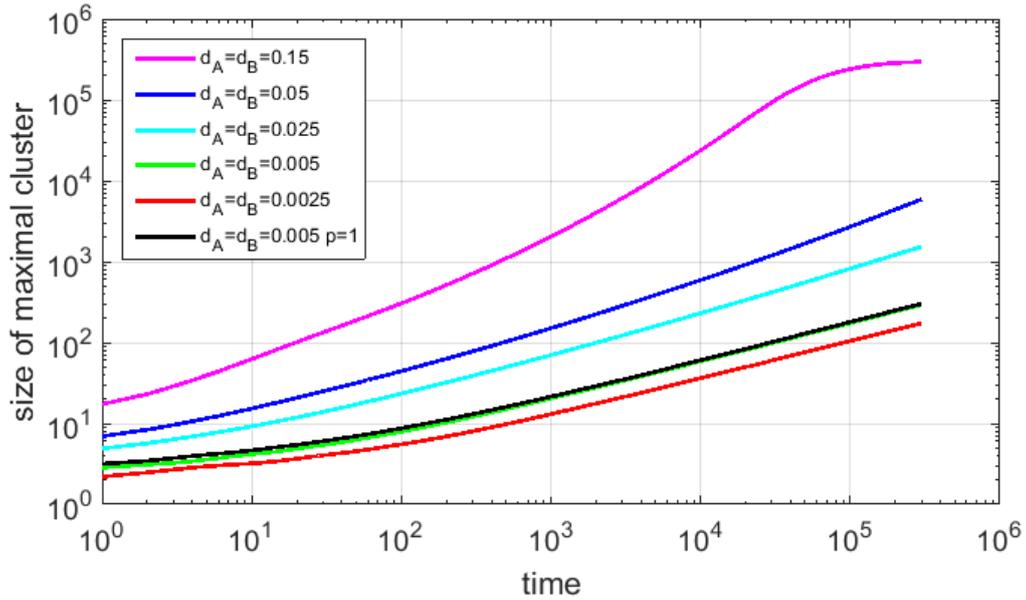

**Supplementary Figure S2**: Mass of the largest cluster vs. time, in the case of a 1000x1000 lattice with periodic boundary conditions, for five droplet concentrations. Average is over 10,000 realizations.

Charge Accumulation

In Supplementary Fig. S3 we plot the absolute value of the total charge $Q = |Q_+ - Q_-|$ on the largest cluster as a function of time, for several different droplet concentrations. We observe that this charge increases with time. In Supplementary Fig. S4 we plot the largest charge found on any cluster as a function of time, and we see that this is also a monotonically increasing function.

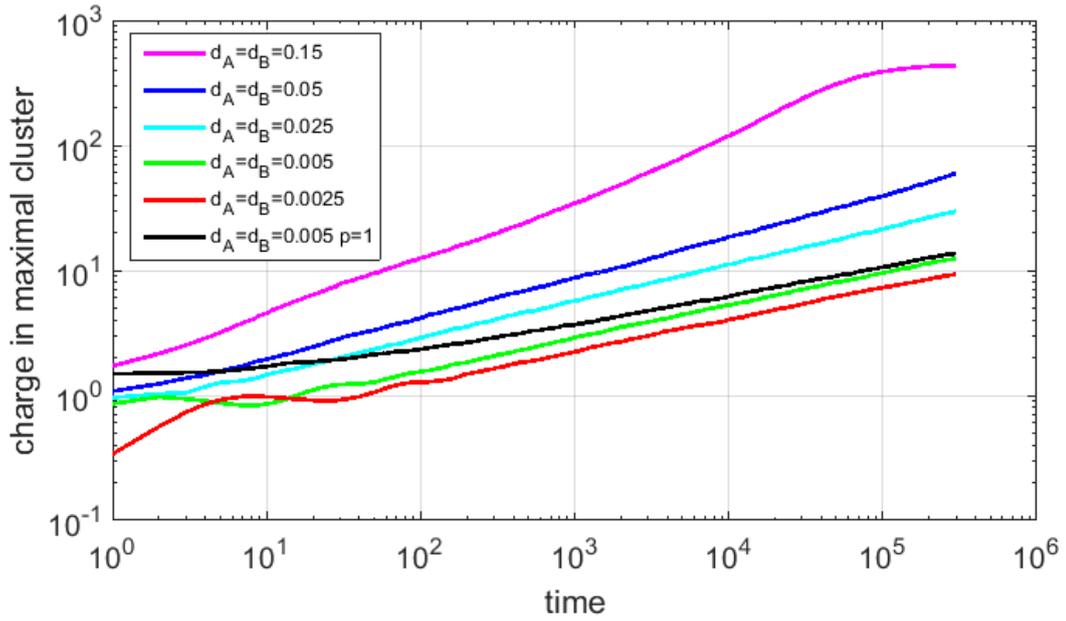

**Supplementary Figure S3**: Charge of the largest cluster vs. time, in the case of a 1000x1000 lattice with periodic boundary conditions. Five droplet concentrations are studied over 300,000 MCS. Average is over 10,000 realizations.

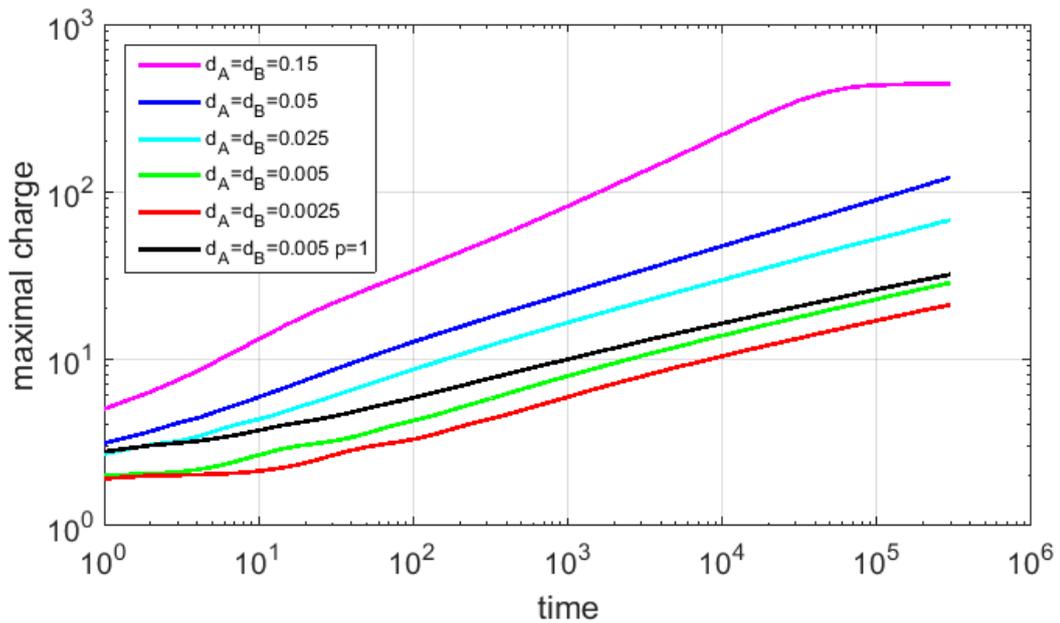

**Supplementary Figure S4:** Maximal charge per single cluster observed on a (not necessarily the largest) cluster vs. time, in the case of a 1000x1000 lattice with periodic boundary conditions. Five droplet concentrations are studied over 300,000 MCS. Average is over 10,000 realizations.

In Supplementary Fig. S5 we plot for several different densities the maximal charge as a function of the size of the cluster that carries this maximal charge and we observe that all densities collapse to the same line with a slope of ~0.5.

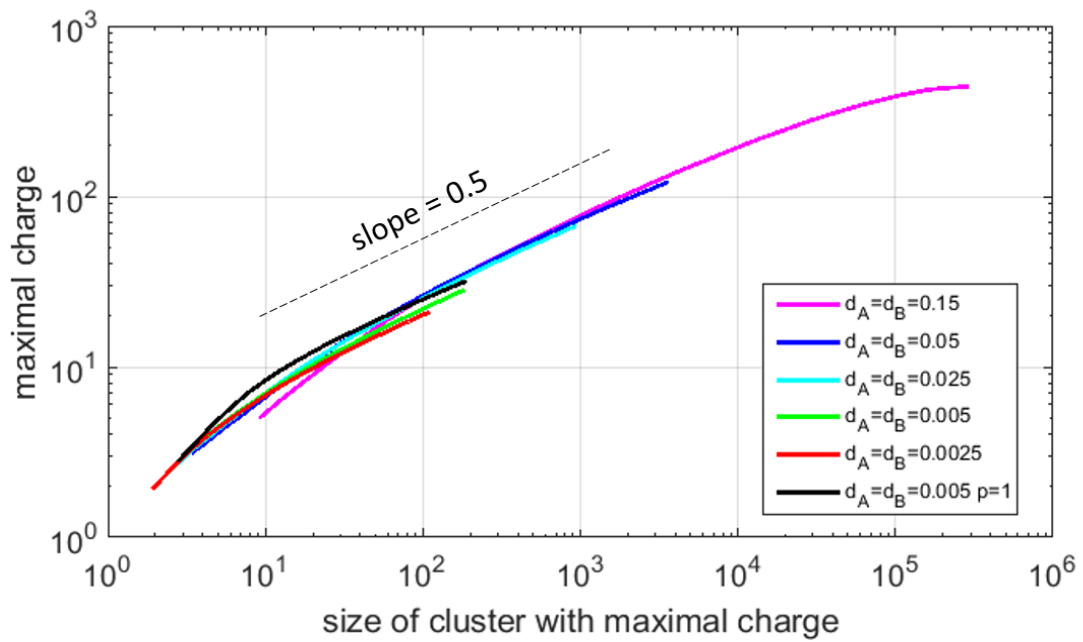

**Supplementary Figure S5:** Maximal charge vs. the mass of the cluster that carries it, in the case of a 1000x1000 lattice with periodic boundary conditions. Five droplet concentrations are studied over 300,000 MCS. Averages are over 10,000 realizations.

In Supplementary Fig. S6 we plot the charge to mass ratio of the cluster carrying the maximal charge, as a function of time. We observe that this ratio decreases with time for all concentrations. The larger the concentration the faster is this decrease.

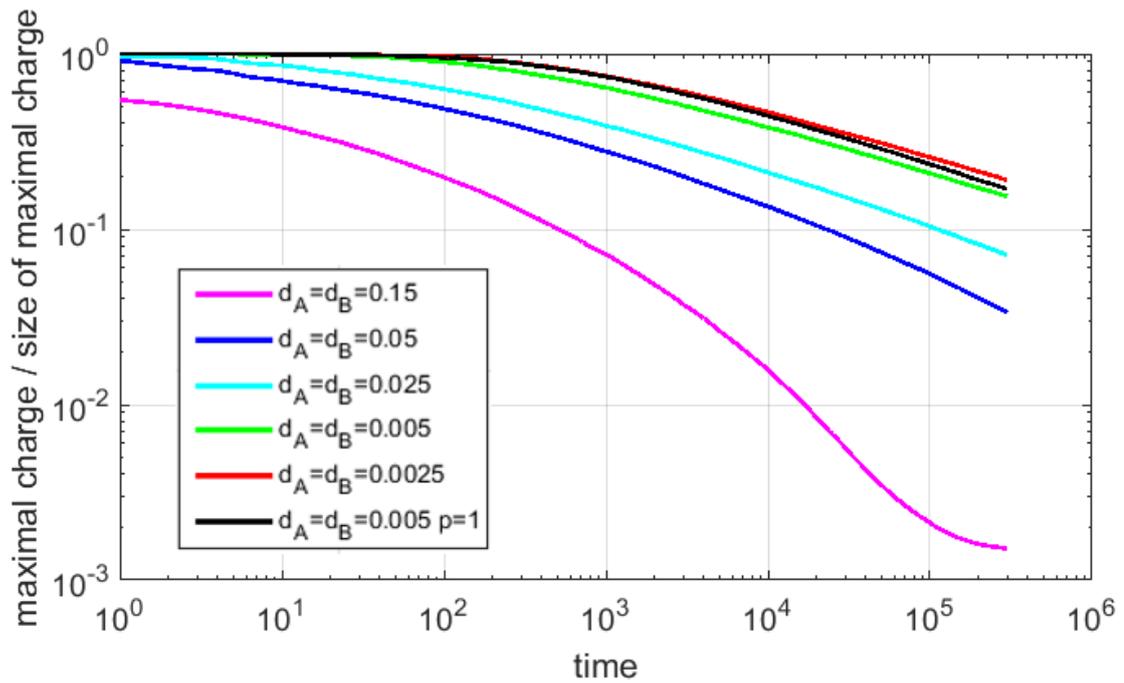

**Supplementary Figure S6:** Maximal charge over the mass of the cluster that holds the maximal charge, plotted vs. time, in the case of a 1000x1000 lattice with periodic boundary conditions. Five droplet concentrations are studied over 300,000 MCS. Averages are over 10,000 realizations.

Like – Unlike Distribution

We now examine the largest cluster (LC) that has been formed, at the longest time period of the simulation. We derive the "like-unlike" distribution, as a measure of the segregation for the positive and negative droplets on the cluster. For each of the droplets of the LC, we count the number of like (i.e. having the same charge) and unlike (i.e. having opposite charges) droplets, that reside at a (Manhattan) distance r from the given droplet. We define $P_{ab}$ and $P_{aa,bb}$ as the number of unlike and like droplets respectively, at a distance r from each droplet of the cluster, divided by the total number of droplets at that distance. The results are shown in Supplementary Fig. S7, where we observe that the difference $P_{ab}$ - $P_{aa,bb}$ decreases with the distance from the original point.

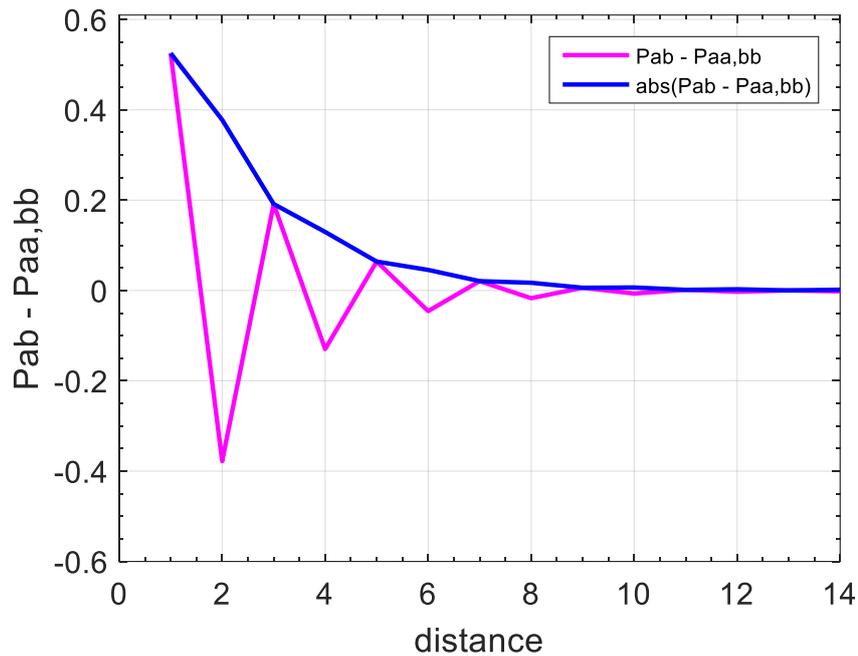

**Supplementary Figure S7**: Difference of like – unlike droplets at a Manhattan distance r, over the total number of droplets at that distance (red line). The absolute difference is shown by the blue line. All results are an average over 100 simulations. Lattice size: 1000x1000 with periodic boundary conditions, concentration $d_A = d_B = 0.05$, merging probability p = 0.1

Distributions of Masses

In Supplementary Fig. S8 we plot the distribution of the cluster masses for the largest cluster at three different time periods, while in Supplementary Fig. S9 we do the same but for all clusters present. All values have been normalized to unity. We observe that at early times sizes of clusters are narrowly distributed, while at later times are more broadly distributed, as expected.

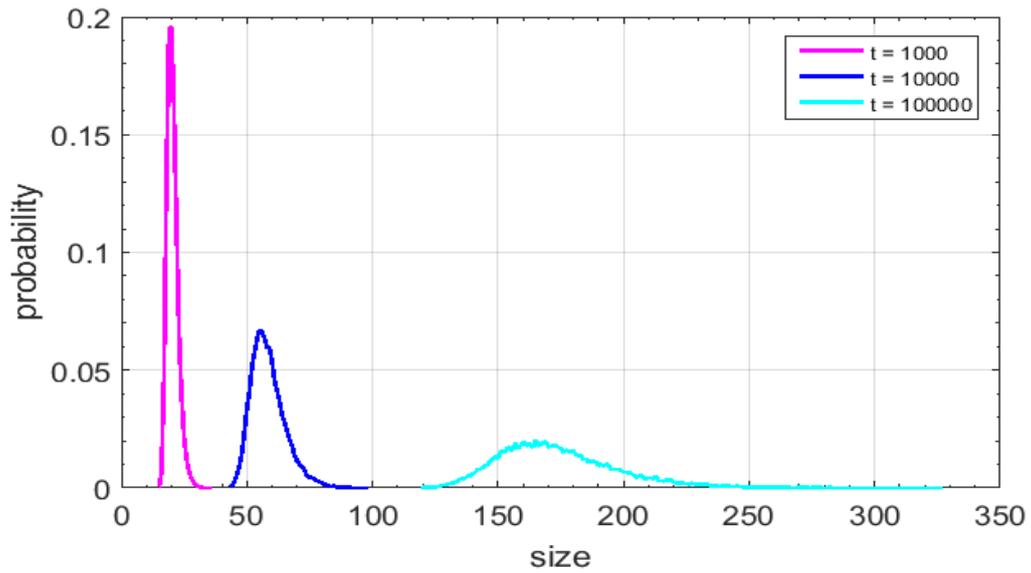

**Supplementary Figure S8:** Mass distribution of maximal-sized cluster, at 3 different time periods. (Average over 20,000 realizations, 1000x1000 lattice) $d_A=d_B=0.005$

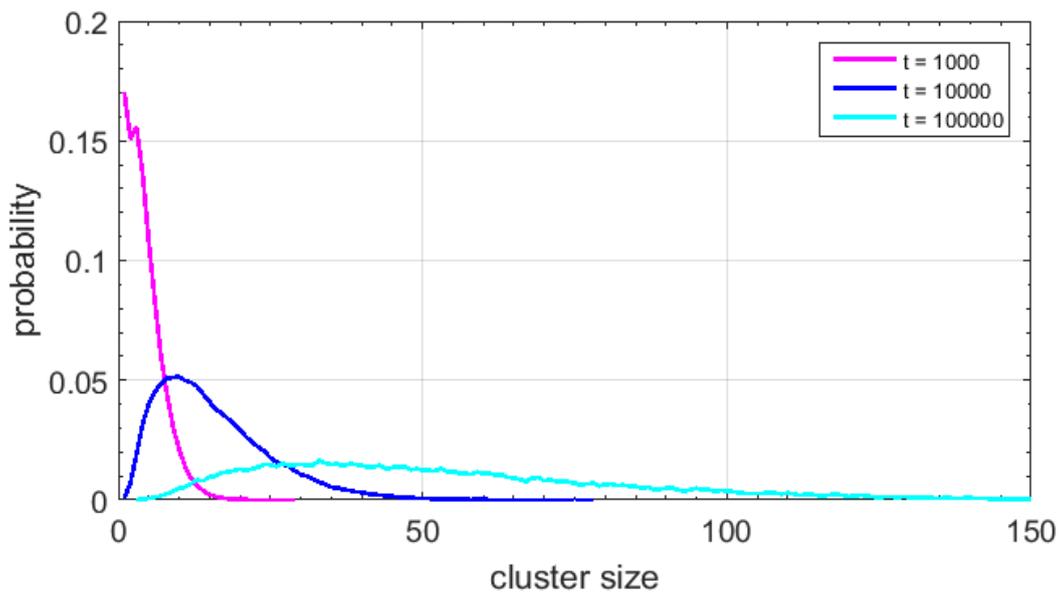

**Supplementary Figure S9:** Mass distribution of all clusters, at 3 different times. (Average over 20,000 realizations, 1000x1000 lattice) $d_A=d_B=0.005$

Electric Dipole Moment

In Supplementary Fig. S10 we plot the dipole moment of the largest cluster as a function of time, which shows that the dipole moment increases with time, as the clusters grow larger. In Supplementary Fig. S11 we plot the largest dipole moment found on any cluster, not necessarily the largest one, as a function of time.

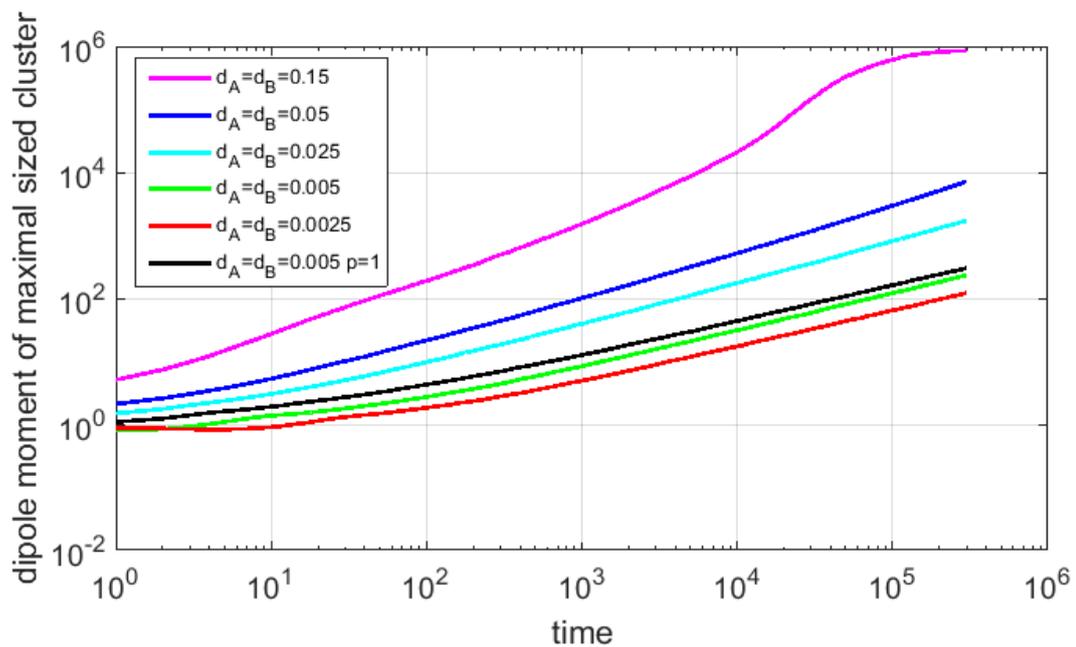

**Supplementary Figure S10:** Dipole moment of the largest cluster vs. time, in the case of a 1000x1000 lattice with periodic boundary conditions. Five droplet concentrations are studied for 300,000 MCS. Average is over 10,000 realizations.

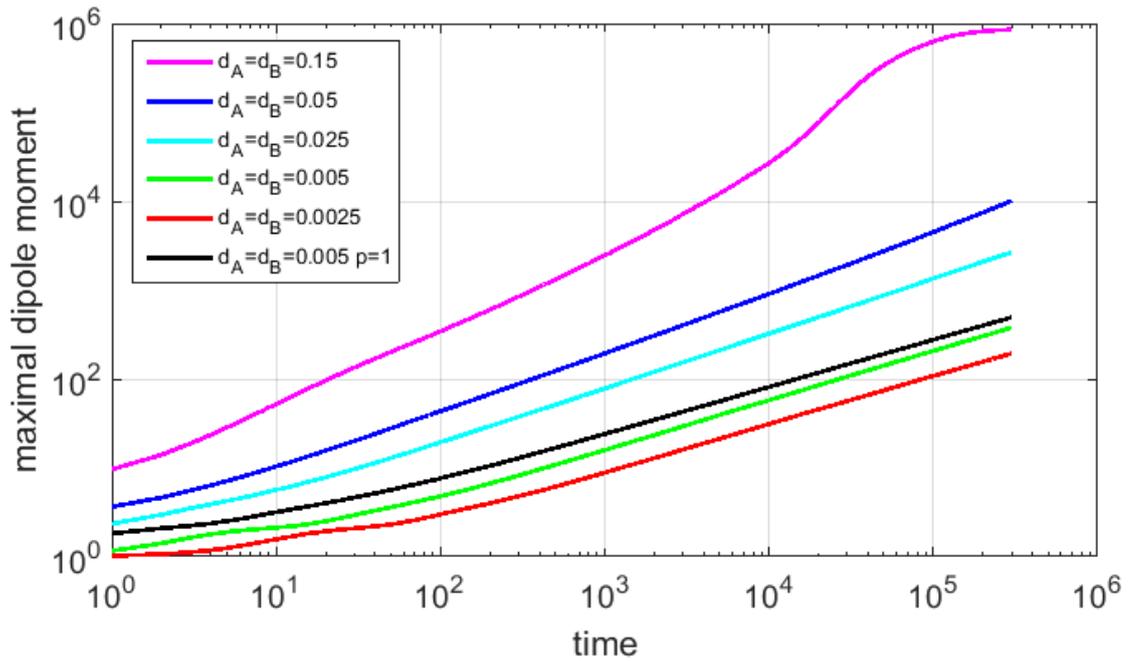

**Supplementary Figure S11:** Max dipole moment of (not necessarily the largest) cluster vs. time, in the case of a 1000x1000 lattice with periodic boundary conditions. Five droplet concentrations are studied for 300,000 MCS. Average is over 10,000 realizations.

Finally, in Supplemental Fig. S12 we plot the maximal dipole moment as a function of the size of the cluster with the maximal dipole moment. We observe a universal behavior for all densities.

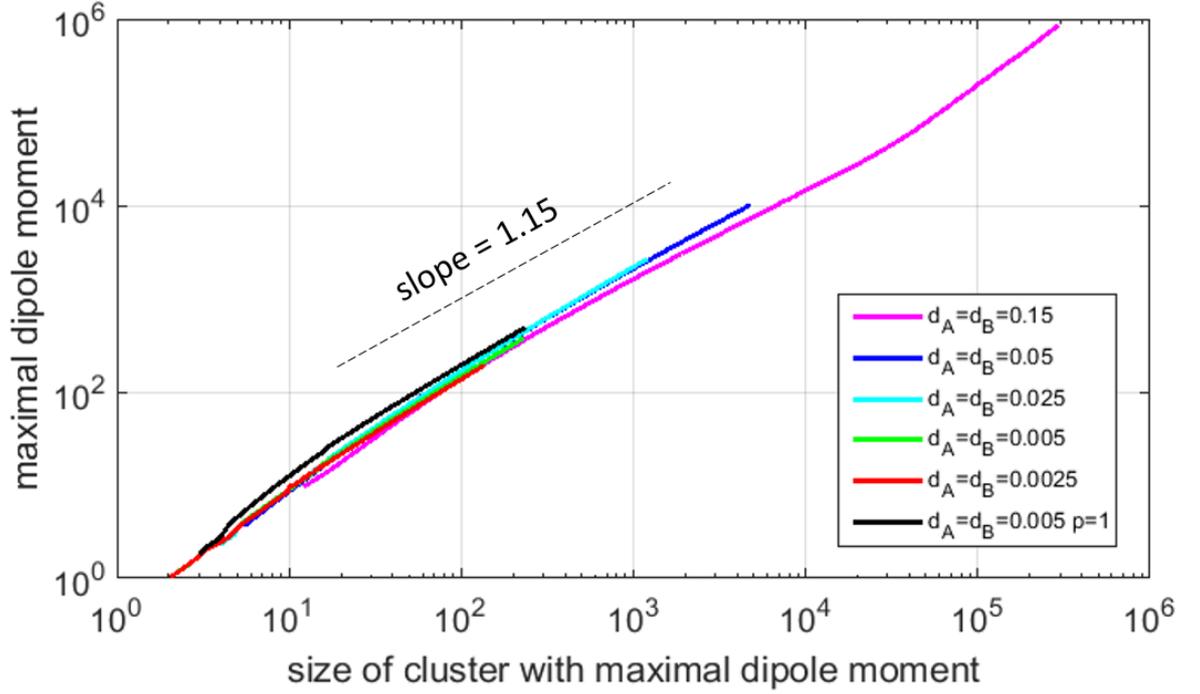

**Supplementary Figure S12:** Max dipole moment of the system vs. its cluster mass, in the case of a 1000x1000 lattice with periodic boundary conditions. Five droplet concentrations are studied for 300,000 MCS. Average is over 10,000 realizations.

Discussion

<u>Two time regimes in Number of clusters evolution</u>

The crossover time between the two regimes is approximately the average time it takes for monomers in the starting configuration to meet each other and start merging into clusters. In the beginning of the simulations we have monomers with total concentration $d_A+d_B = d$ ($0<d<1$), so we have on average 1 droplet in a square box of $1/d$ lattice sites. Such a box has edges with length $(1/d)^{1/2}$, so the average distance between two monomers at the starting configuration is $(1/d)^{1/2}$.

The monomers diffuse by a random walk, so the time they need to cover that distance is:

$$t_d = \left(\sqrt{\frac{1}{d}}\right)^2 = \frac{1}{d} \qquad (S1)$$

For the concentrations we have used in our simulations, these crossover times are 3.3, 10, 20, 100 and 200 steps which are in agreement with Supplementary Fig. S1, as shown with the dotted vertical lines.

Electric Dipole Moment

The dipole moment $\vec{p}$ of N charged particles and non-zero overall (total) charge is defined as:

$$\vec{p} = \sum (\vec{r}_i - \vec{r}_{ref})\, q_i \qquad (S2)$$

where the sum runs over all charged particles of the cluster; the vector $\vec{r}_i$ is the position of the *i*th particle, <u>measured from an arbitrary origin</u>; $q_i$ is its charge and $\vec{r}_{ref}$ a reference point. For non-neutral systems - like the clusters in the case of this study - the value of $\vec{p}$ given by Supplementary Eq. (S2) depends on the point of reference, thus rendering it useless in the case of a moving cluster. To overcome this, the point of reference must be a point on the cluster. The conventional choice of such a reference point is the *center of mass,* given by:

$$\vec{r}_{cm} = \frac{\sum \vec{r}_i\, m_i}{\sum m_i} \qquad (S3)$$

Therefore, we calculate the dipole moment given by:

$$\vec{p}_{cm} = \sum (\vec{r}_i - \vec{r}_{cm})\, q_i \qquad (S4)$$

According to Supplementary Eq. (S4), the calculation of the dipole moment of a cluster at a given time calls for summation over all its charged particles. This calculation is significantly simplified if we take advantage of the fact that, in the case studied (i.e. the particles have identical masses and absolute charges) the center of mass and center of charge of the positively (negatively) charged particles are one and the same. Therefore, we can replace the distribution of the positive (negative) particles by a body with mass and charge equal to the total mass and charge of the positive (negative) particles, placed on the center of mass/charge of these particles.

Thus, we end up with a "dipole" having the mass and charge of the positive and negative particles concentrated on its positive and negative ends, respectively. Now, the calculation of the dipole moment of the cluster is reduced to calculating the dipole moment of this "dipole" with respect to its center of mass, which is identical to the center of mass of the whole cluster (Supplementary Fig. S13). The dipole moment magnitude of this dipole is given by:

$$p_{cm} = |q_A| \cdot r_{cmA} + |q_B| \cdot r_{cmB} \quad (S5)$$

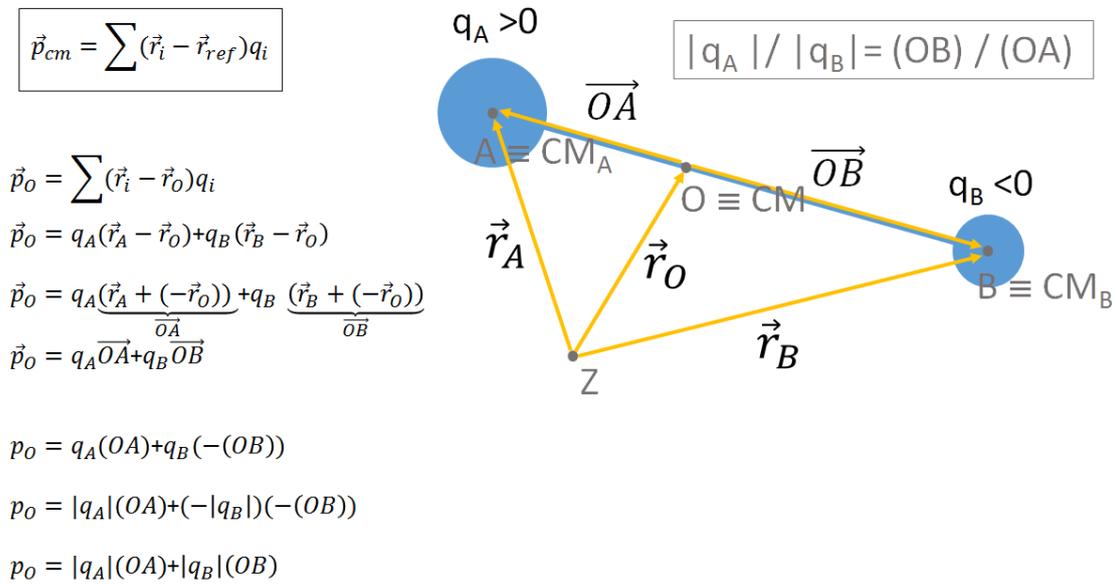

Supplementary Figure S13: Dipole moment of the dipole having the mass and charge of the positive and negative cluster particles accumulated on its positive and negative ends, respectively. The two blue spheres have different sizes, indicating that they could represent different amounts of charge.

We differentiate between two distinct cases: (a) translational cluster movement without merging and (b) translational cluster movement with merging. The dipole moment of a cluster, given by Supplementary Eq. (S4), does not change with the motion of the cluster. Consequently, in case (a), it suffices to update the position of the center of mass of the positive, negative and total particles, $\vec{r}_{cmPos}$, $\vec{r}_{cmNeg}$ and $\vec{r}_{cm}$, respectively, in order to keep track of their positions. In case (b), we calculate the $\vec{r}_{cmPos}$, $\vec{r}_{cmNeg}$ and $\vec{r}_{cm}$ of the resulting merged cluster from the respective properties of its merged cluster components, and then use Supplementary Eq. (S5) to derive the new value of the dipole moment.

In a system with periodic boundary conditions, two particles situated near two opposing boundaries of the system are considered to be in the vicinity of one another. As a consequence, using Supplementary Eq. (S4) and (S5) to calculate the center of mass would produce incorrect results. A better suited approach in this case is to treat each coordinate, *x* and *y*, as if it were on a circle instead of a line. The procedure is described in detail in[18].

The results of the simulations indicate that the relation between the dipole moment P and the mass N of the cluster is (see Fig. 3):

$$P \sim N^a \qquad (S6)$$

where the exponent α is in the range 1.07 – 1.24.

This result is rationalized as follows: N (positive and negative) is the number of droplets ($N_+ \approx N_- \approx N/2$), i.e. the mass of the cluster. **R** is the cluster's linear dimension, and **$X_i$** is the position (abscissa) of droplet *i*, on the X axis (*i* = 1 ... N). Focusing on the **horizontal axis**, we make the following assumption: The abscissas of the droplets are uniformly distributed in [0, R] and independent. Therefore, the position $X_i$ can be seen as a random variable, assuming independent random values, uniformly distributed in the range [0, R] (see Fig. 5)

The mean, variance and standard deviation of such a random variable, are:

$$\bar{X}_i = \frac{R}{2} \qquad (S7a)$$

$$\sigma^2_{X_i} = \frac{R^2}{12} \qquad (s7b)$$

$$\sigma_{X_i} = \frac{R}{\sqrt{12}} \qquad (S7c)$$

If we take a sample of the random variables $X_i$, the mean μ (i.e. the center of mass), the variance of the mean $\sigma^2_{cm}$, and the standard deviation $\sigma_{cm}$ of the mean are:

$$\mu = \sum \frac{X_i}{N} = \frac{R}{2} \qquad (S8a)$$

$$\sigma^2_{cm} = \frac{\sigma_{X_i}^2}{N} \qquad (S8b)$$

$$\sigma_{cm} = \frac{\sigma_{X_i}}{\sqrt{N}} \qquad (S8c)$$

The mass of the cluster depends on its linear dimension R, following a power law:

$$N \sim R^{d_f}$$

where $d_f$ is its fractal dimension. It follows that:

$$R \sim N^{1/d_f} \qquad (S9)$$

Combining (S7c), (S8c) and (S9) we derive the following relation:

$$\sigma_{cm} \sim \frac{N^{1/d_f}}{\sqrt{12N}} \sim N^{\frac{1}{d_f} - \frac{1}{2}} \qquad (S10)$$

We will use Supplementary Eq. (S10), to find an expression for the relation between the dipole moment P of the cluster and its mass N (mass).

The total positive and negative charges are almost equal, and of the order of N:

$$Q \approx Q_+ \approx Q_- \approx N/2 \qquad (S11)$$

The dipole moment P is defined as the product of the charge Q, times the distance Δx of the centers of mass of the positive and negative charges. The distance Δx is of the order of $\sigma_{cm}$:

$$\Delta x \sim \sigma_{cm} \qquad (S12)$$

By combining (S10) and (S12), we get:

$$\Delta x \sim N^{\frac{1}{d_f} - \frac{1}{2}} \qquad (S13)$$

Since the same reasoning holds for the y-axis, we can assume that $\Delta y \sim \sigma_{cm}$, and

$$\Delta r = \sqrt{\Delta x^2 + \Delta y^2} \sim \sigma_{cm}$$

is also of the order of $\sigma_{cm}$.

Finally, from (S11) and (S13) we get for the dipole moment P:

$$P \sim Q \cdot \Delta r \qquad (S14)$$

$$P \sim N^{\frac{1}{2} + \frac{1}{d_f}} \qquad (S15)$$

Therefore, the exponent α is:

$$\alpha = \frac{1}{d_f} + \frac{1}{2} \qquad (16)$$